\documentclass[epj,twocolumn]{webofc}
\usepackage[varg]{txfonts}   
\usepackage{subfigure}
\usepackage{subfig}

\woctitle{17th International Symposium on Very High Energy Cosmic Ray Interactions (ISVHECRI 2012), Berlin, Germany, 2012}
\begin{document}

\title{The LHCb experiment: status and recent results}
%
%

\author{Dmytro Volyanskyy \small{(on behalf of the LHCb collaboration)}}

\institute{Max-Planck-Institut f\"ur Kernphysik, PO Box 103980, 69029 Heidelberg, Germany}

\abstract{%
The LHCb experiment is one of the major research projects at the Large Hadron Collider.
Its acceptance and instrumentation is optimised
to perform high-precision studies of flavour physics and particle production
in a unique kinematic range at unprecedented collision energies.
Using large data samples accumulated in the years 2010--2012, 
the LHCb collaboration has conducted a series of measurements
providing a sensitive test of the Standard Model and strengthening 
our knowledge of flavour physics, QCD and electroweak processes.
The status of the experiment and some of its recent results
are presented here.
}
\maketitle

 
\section{The LHCb experiment}
\label{intro}
 
Due to its unprecedented interaction rate and collision energy, 
the Large Hadron Collider~(LHC) delivers 
a vast amount of heavy flavour particles. 
This provides an excellent opportunity to perform 
high-precision measurements and explore rare processes 
in the heavy flavour sector. The Large Hadron Collider beauty~(LHCb) 
experiment has been mainly designed to study the physics 
of the heaviest hadrons - the beauty flavoured ones~($B$ hadrons).
In particular, the experiment aims to explore 
$CP$-violating phenomena in the $B$ hadron sector 
and highly suppressed flavour-changing neutral currents~(FCNC) 
using appropriate $B$ hadron decay channels plus
overconstrain the CKM unitary triangle.
These measurements should provide a rigorous 
test of the Standard Model while indirectly probing for New Physics effects.
The latter may appear as contributions from new virtual heavy particles 
in loop-mediated processes giving access to scales greater 
than the LHC centre-of-mass energy. 
Another important goal of the experiment is to fulfil 
a broad charm physics programme.
\begin{figure}[b!]
\centering
\resizebox{3.2in}{!}{
\rotatebox{0}{
\includegraphics{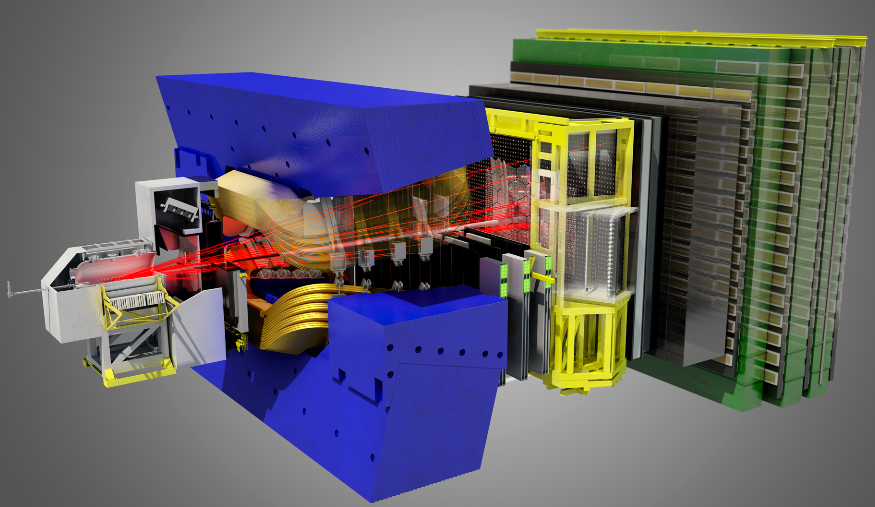}}}
\caption{View of the LHCb spectrometer.}
\label{fig:LHCb} 
\end{figure}

Since $B$ hadrons at the LHC are mainly 
produced at low polar angle in the same forward cone, 
the LHCb detector~\cite{lhcb} has been constructed as a single-arm, forward spectrometer, 
as can be seen in Fig.~\ref{fig:LHCb}.
Its angular coverage ranges from 10~mrad to 300~(250)~mrad in the bending~(non-bending) plane
allowing particle reconstruction in the pseudorapidity range $2<\eta<5$.
Although the spectrometer covers just about $4\%$ of the solid angle, 
it detects about $40\%$ of heavy quark hadrons produced at the LHC. 
Due to its unique pseudorapidity coverage and the ability to perform
measurements at low transverse momenta $p_{\rm T}$, the detector
also permits a unique insight into particle production in the forward
region at the LHC.

\section{Detector performance and data taking in 2009--2012}
\label{performance}

\begin{figure}[t!]
\centering
\resizebox{3.2in}{!}{
\includegraphics{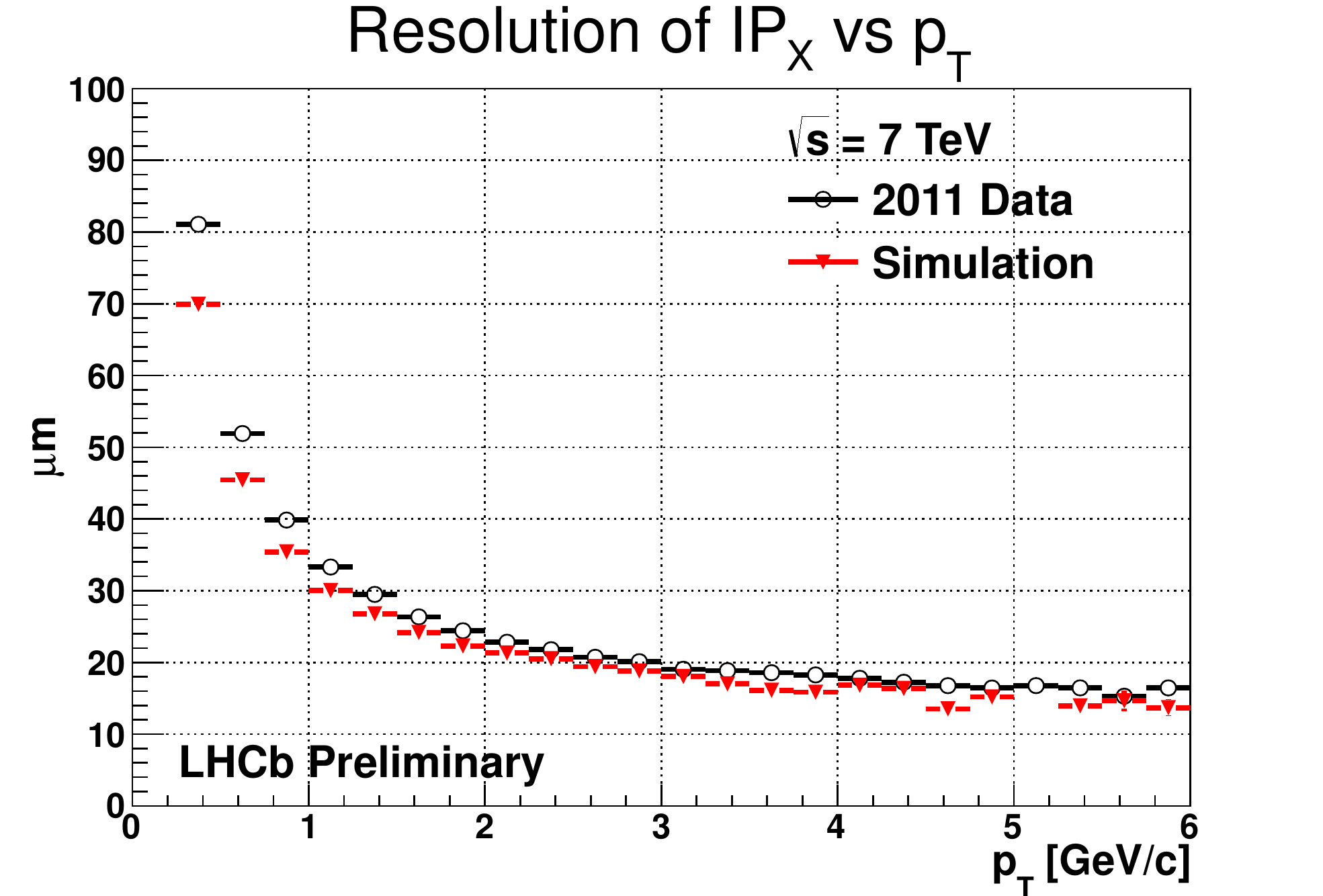}}
\caption{Impact parameter resolution along the X-axis as a function $p_{\rm T}$.}
\label{fig:velo3} 
\end{figure}

The LHCb subcomponents can be split into the following two categories: 
tracking detectors, which are used to reconstruct the trajectories 
and momenta of charged particles, and particle identification detectors, 
which are employed to identify different types of particles.
The LHCb tracking system covers the full acceptance of the experiment which is unique at the LHC.  
It consists of a silicon-strip vertex detector~(VELO) 
surrounding the collision region, a large-area silicon-strip detector located 
upstream of a dipole magnet with a bending power of about $4{\rm\,Tm}$, 
and three stations of silicon-strip detectors and straw drift tubes placed downstream. 
The VELO has a larger angular acceptance than the rest of the spectrometer, including partial 
coverage of the backward region which allows reconstruction of charged particle tracks 
in the pseudorapidity ranges $1.5<\eta<5.0$ and $-4<\eta<-1.5$.
Its sensors are positioned along and perpendicular to the beam axis being 
split into two movable halves. Their sensitive area during stable beam conditions starts 
at 8~mm from the beam axis, providing full azimuthal coverage. 
The VELO measures the radial and azimuthal track coordinates 
\begin{figure}[t!]
\centering
\resizebox{3.2in}{!}{
\includegraphics{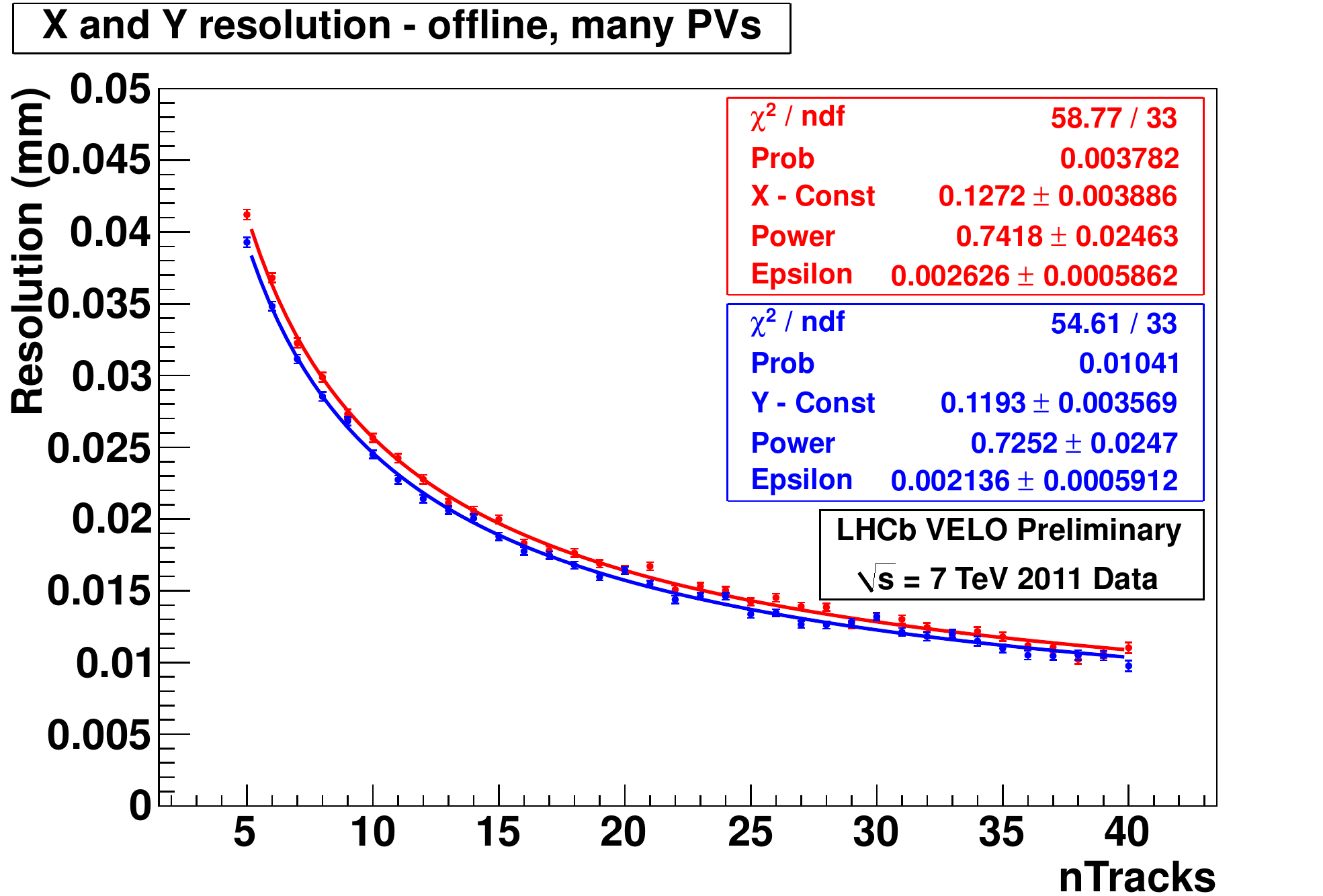}}
\newline
\resizebox{3.1in}{!}{
\includegraphics{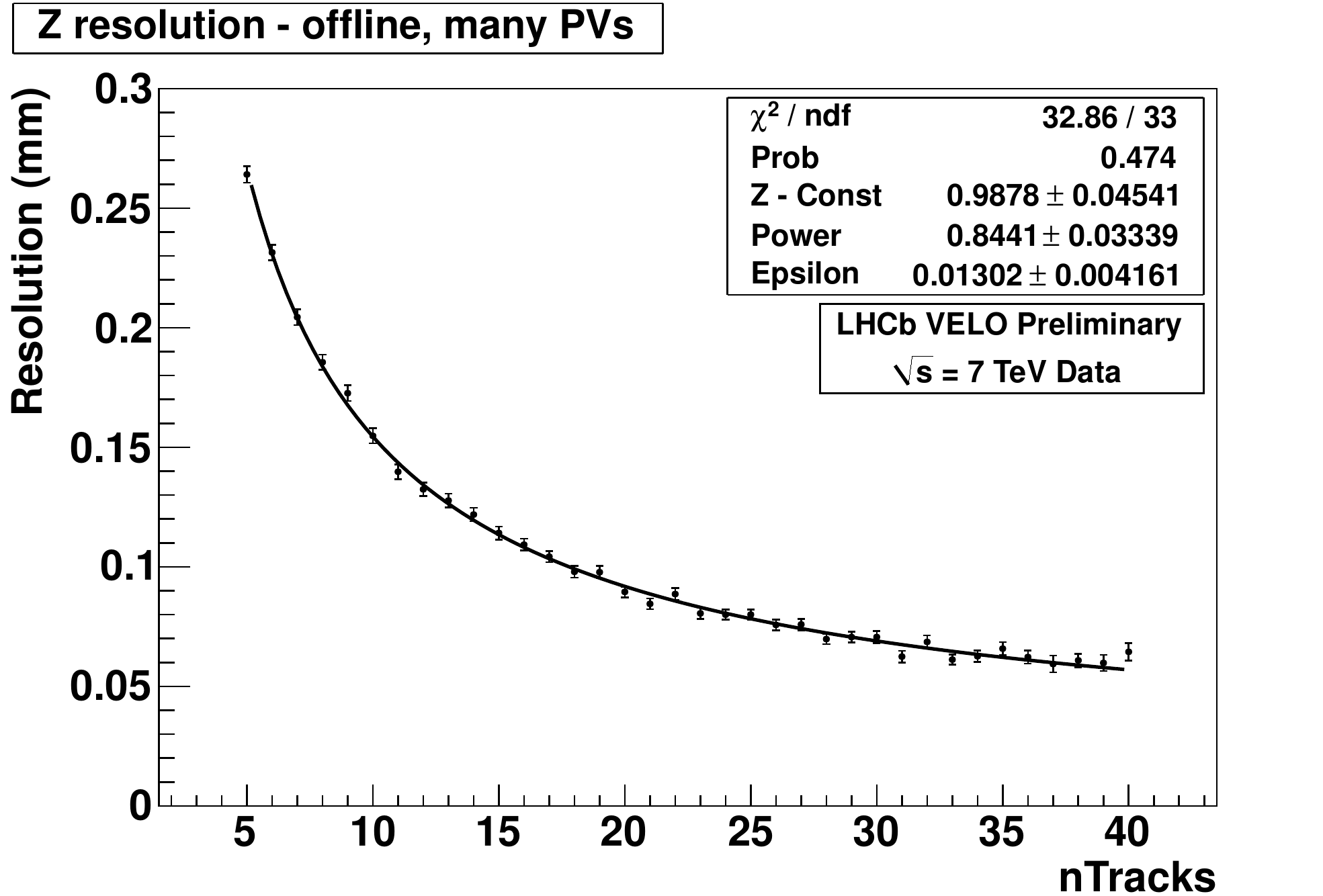}}
\caption{Primary vertex resolution along the X~(shown in red), Y~(shown in blue) and Z~(shown in black) axes as a function of the number of tracks used in the vertex fit.}
\label{fig:velo12} 
\end{figure}
around the luminous region with high precision. 
In particular, a hit resolution of 4~$\mu$m is achieved at the minimal pitch 
and optimal track angle along with a hit finding efficiency over $99\%$ and 
an alignment accuracy of 1~$\mu$m. This allows high-precision reconstruction 
of the positions of the primary and decay vertices. 
Figure~\ref{fig:velo12} illustrates the achieved primary vertex resolution 
as a function of the number of tracks used in the vertex fit,  
while the impact parameter resolution along the X-axis is shown in Fig.~\ref{fig:velo3} as a function of $p_{\rm T}$.
\begin{figure}[th!]
\centering
\resizebox{3.2in}{!}{
\includegraphics{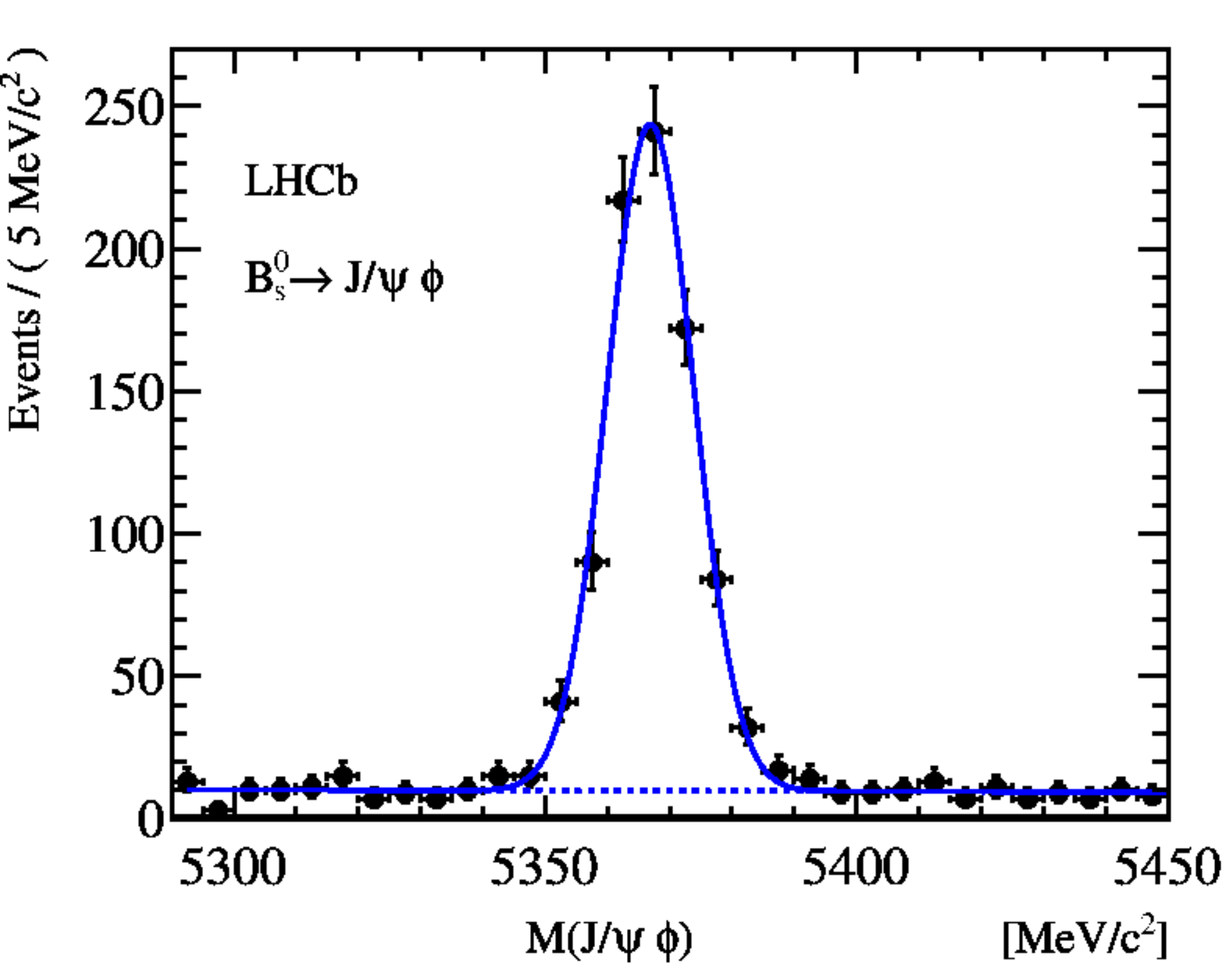}}
\newline
\resizebox{3.2in}{!}{
\includegraphics{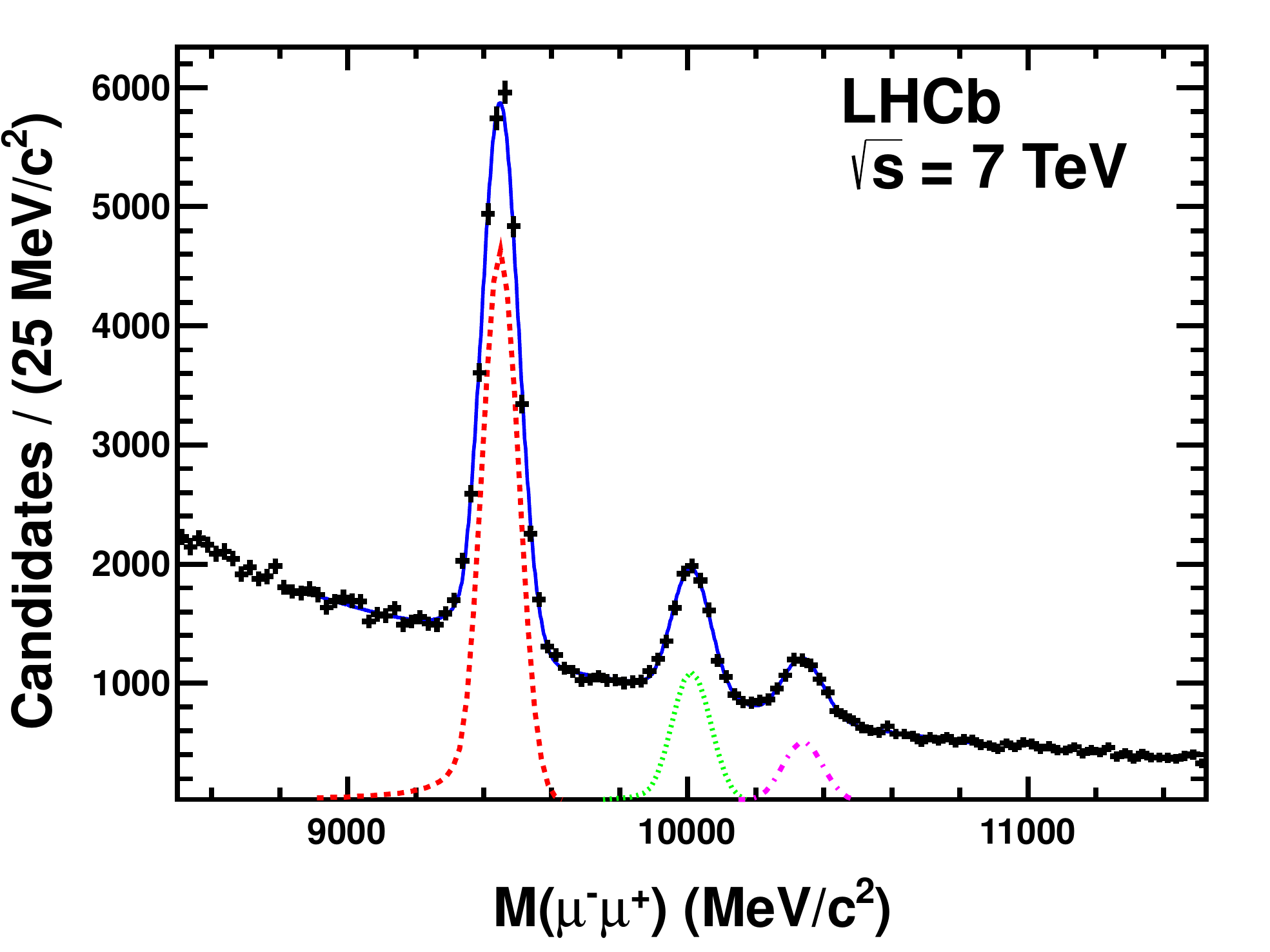}}
\caption{Invariant mass distribution of the $B^{0}_{s}$ meson decaying into $J/\psi\phi$ resonances (top) and $\Upsilon$ mesons decaying into two muons~(bottom). 
         The corresponding mass resolutions are 7~MeV/$c^{2}$ for the $B^{0}_{s}$ meson and 54~MeV/$c^{2}$ for the $\Upsilon(1S)$ meson.}
\label{fig:mass1} 
\end{figure}
\begin{figure}[t!]
\centering
\resizebox{3.2in}{!}{
\includegraphics{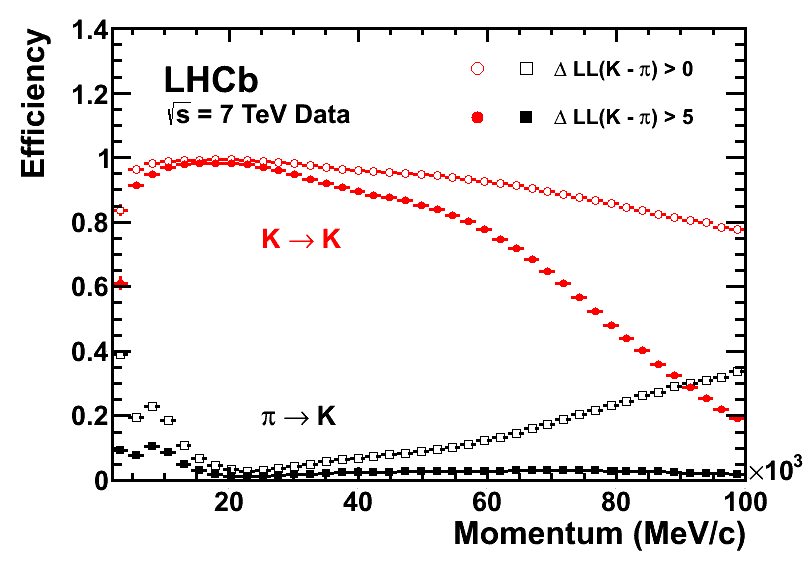}}
\newline
\resizebox{3.2in}{!}{
\includegraphics{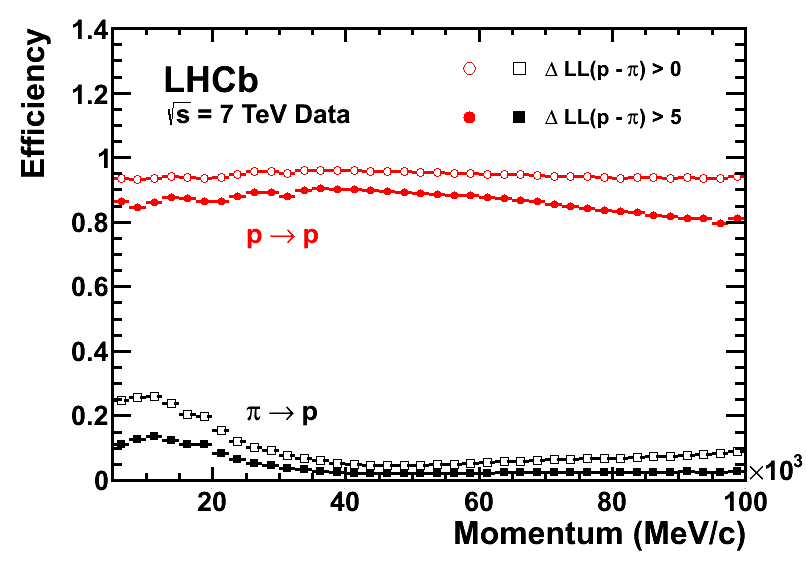}}
\caption{Kaon~(top) and proton~(bottom) identification efficiency~(circles) and pion misidentification rate~(squares) 
         as a function of momentum for tight and loose selections.
        }
\label{fig:rich} 
\end{figure}
The combined tracking system has a momentum resolution $\Delta p/p$ that varies 
from $0.4\%$ at 5~GeV/$c$ to $0.6\%$ at 100~GeV/$c$, and an impact parameter resolution 
of about $20~\mu$m for tracks with high $p_{\rm T}$. The overall track finding efficiency 
is found to exceed $96\%$ for tracks with momentum greater than 5~GeV/$c$.

An excellent performance of the LHCb tracking system 
permits high-precision measurements of the invariant mass 
of different resonances decaying into charged particles.~In particular, 
the LHCb collaboration has performed the world's most precise 
measurement of several $B$ hadron masses, 
which was achieved with just $35~{\rm pb}^{-1}$ 
of data collected in 2010~\cite{Bmass}. Examples of different invariant mass distributions 
are shown in Fig.~\ref{fig:mass1}.

Due to the excellent vertexing capabilities provided by the VELO, 
the LHCb spectrometer measures the proper decay time of 
$B$ hadrons with resolution of better than 50~fs. 
This allows a precise resolving of rapid $B^{0}_{s}-\bar{B}^{0}_{s}$ 
oscillations which is essential for time-dependent mixing and $CP$ violation studies in this system.

Robust particle identification plays an important role in 
the suppression of combinatorial background, 
separation of decay channels with identical topology 
and flavour tagging. Charged hadrons at LHCb are identified using 
two ring-imaging Cherenkov detectors with three different radiators. 
These provide efficient $K/\pi$ and $p/\pi$ separations 
over the momentum range $2<p<100~{\rm GeV}/c$.
Figure~\ref{fig:rich} illustrates the efficiency of kaon and proton identification 
along with the corresponding pion misidentification rate as a function of momentum 
for tight and loose selection criteria. For the majority of kaons 
originating from $B$ hadron decays the identification efficiency 
is found to be above $95\%$ with a pion misidentification rate below $10\%$. 

\begin{figure}[t!]
\centering
\resizebox{3.2in}{!}{
\includegraphics{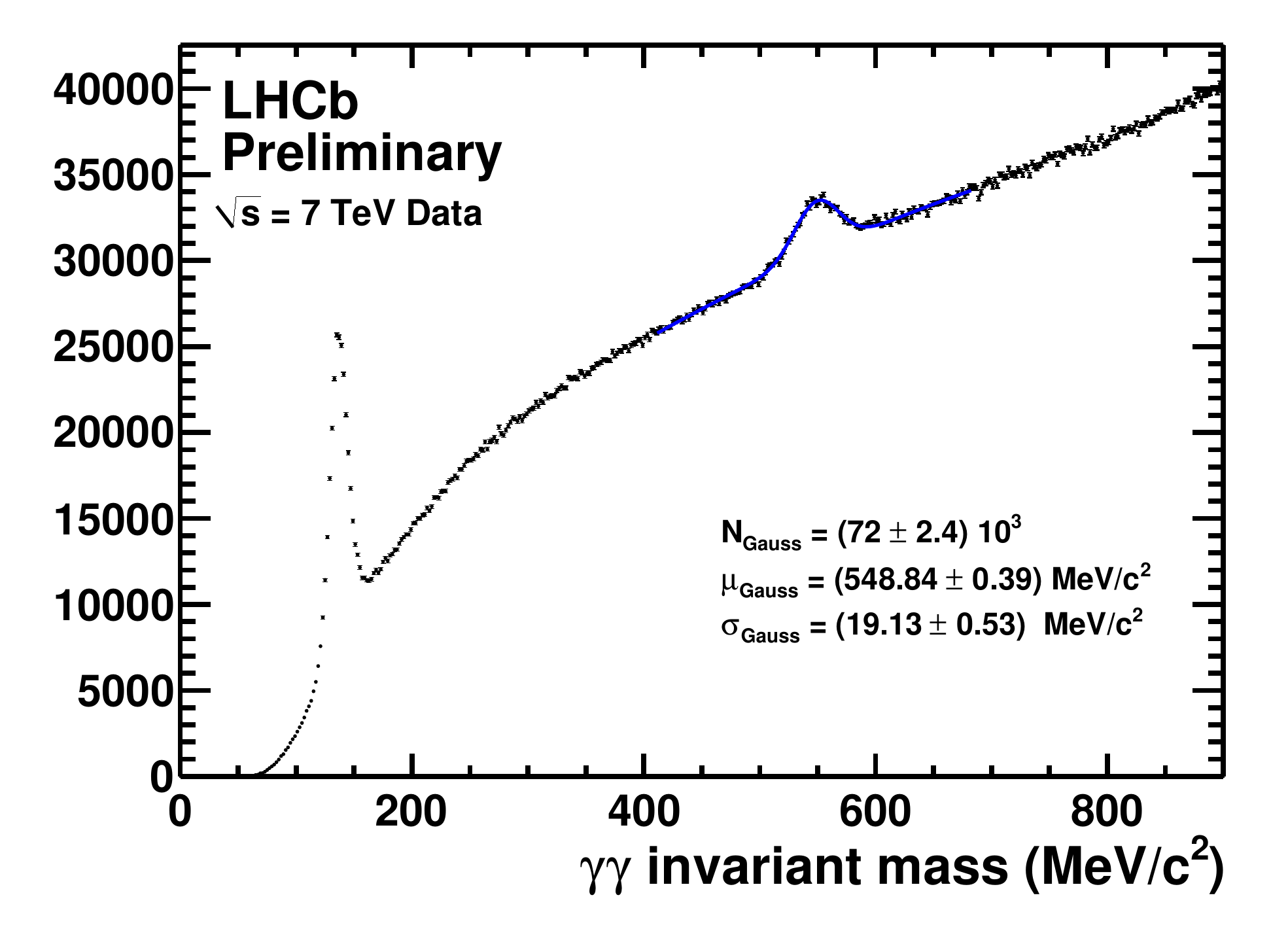}}
\newline
\resizebox{3.2in}{!}{
\includegraphics{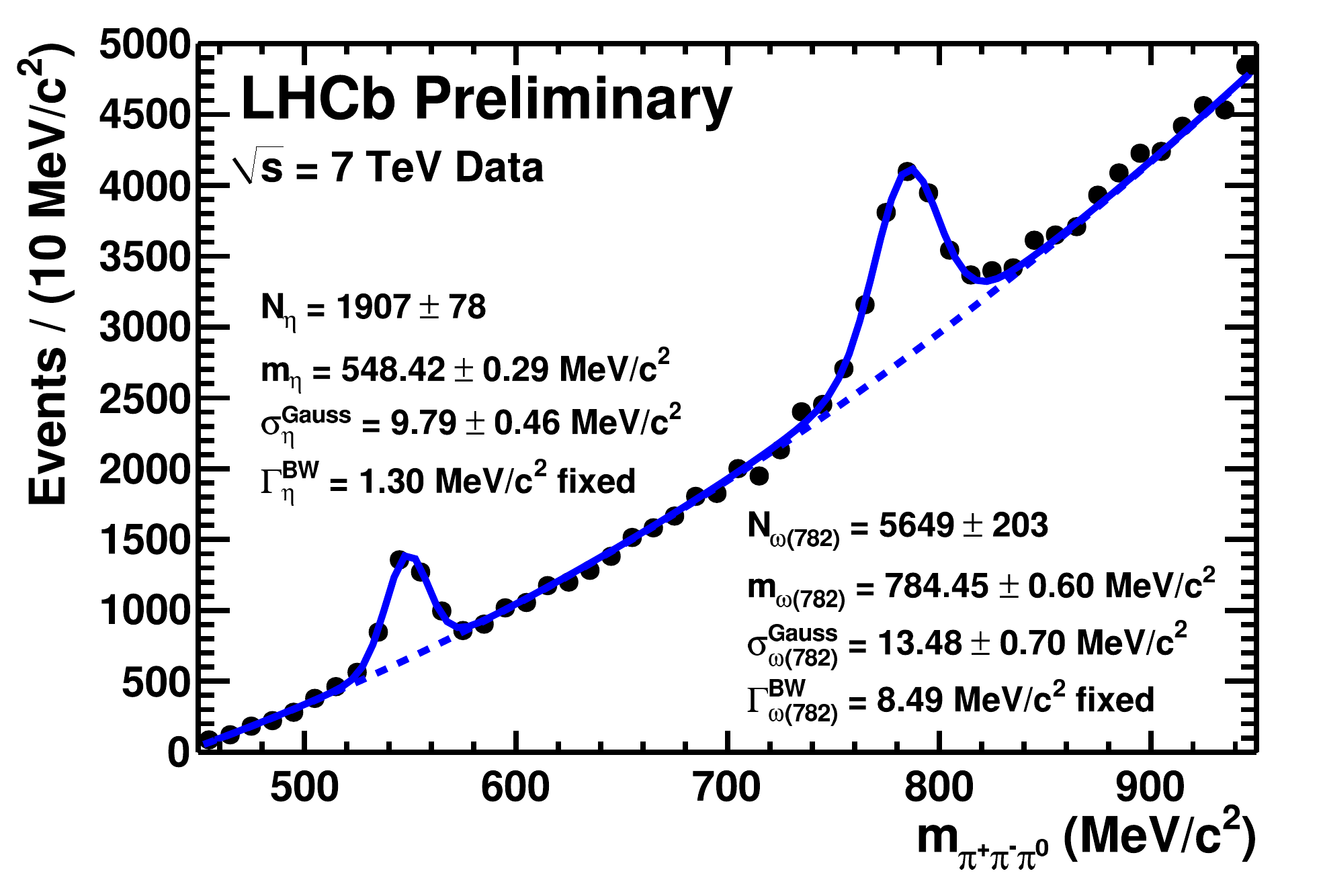}}
\caption{Diphoton invariant mass distribution~(top) with $\pi^{0}$ plus $\eta$ peaks 
        and $\pi^{+}\pi^{-}\pi^{0}(\gamma\gamma)$ invariant mass distribution~(bottom) 
        with $\eta$ plus $\omega(782)$ peaks. }
\label{fig:ecal} 
\end{figure}

Photon, electron and hadron candidates are distinguished 
by a calorimeter system consisting of scintillating-pad and preshower detectors, 
an electromagnetic calorimeter~(ECAL) and a hadronic calorimeter~(HCAL). 
Electrons and photons are fully absorbed in the ECAL which measures their energies 
with a resolution of $\sigma(E)/E=10\%/\sqrt{E}\oplus 1\%$ (with $E$ in GeV). 
This allows reconstruction of various decay channels involving these particles 
as illustrated in Fig.~\ref{fig:ecal}.
The energy resolution of the HCAL is $\sigma(E)/E=69\%/\sqrt{E}\oplus 9\%$. 
In contrast to the ECAL, its information is not directly used 
in the offline analyses, while being very important for triggering purposes.

Muon detection is performed by a system positioned downstream 
of the calorimeters and composed of alternating layers of iron and multiwire proportional chambers. 
The overall muon identification efficiency 
exceeds $95\%$ with a misidentification rate 
of about $3\%$ which is mainly due to decays in flight.

The LHCb experiment possesses a selective and flexible trigger system which allows 
the recording of events with different topologies. It consists of a hardware stage, 
based on information from the calorimeters and muon systems, 
followed by a software stage which applies a full event reconstruction.
The maximum output rate of the hardware and software stages 
is currently limited to about 1~MHz and 5~kHz, respectively.

First proton-proton~($pp$) collisions at 
the LHCb interaction point occurred in November 2009.
Since then and until the end of 2012, the experiment has recorded 
around $3~{\rm fb}^{-1}$ of $pp$ collision data 
at four different centre-of-mass energies~($\sqrt{s}$) 
as summarised in Table~\ref{tab:data}.~Around $99\%$ of 
read-out channels were operational 
throughout the entire data taking period and 
only $1\%$ of accumulated data was discarded for physics analyses. 
To expand its physics programme, the experiment also participates in the proton-lead collision data taking.
\begin{table}[t!]
\centering
\begin{tabular}{c|c|c}
Year & $\sqrt{s}$ (TeV) & Data \\
\hline
2009 &  0.9  &  6.8~$\mu$${\rm b}^{-1}$  \\
2010 &  0.9  &  0.3~${\rm nb}^{-1}$      \\
2010 &  7.0  &  37~${\rm pb}^{-1}$       \\
2011 &  2.76 &  0.1~${\rm pb}^{-1}$      \\
2011 &  7.0  &  1.0~${\rm fb}^{-1}$      \\
2012 &  8.0  &  2.0~${\rm fb}^{-1}$      \\
\end{tabular}
\caption{$pp$ collision data recorded in 2009--2012.}
\label{tab:data}
\end{table}
\begin{figure}[th!]
\centering
\resizebox{3.2in}{!}{
\rotatebox{0}{
\includegraphics{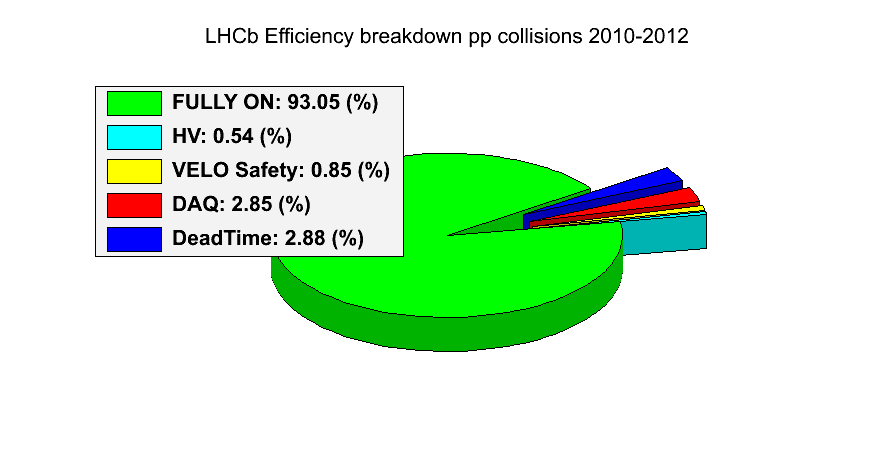}}}
\caption{Data taking efficiency breakdown in 2010--2012.}
\label{fig:LHCbEff} 
\end{figure}

The LHCb detector was originally designed to run at moderate luminosity as 
high pile-up environment complicates flavour tagging and identification of 
$B$ hadron decay vertices. Nevertheless, the majority of the data in 
2011 and 2012 were recorded at an instantaneous luminosity nearly a factor of 
two above the LHCb design value ($L=2.0 \times 10^{32}~{\rm cm^{-2} s^{-1}}$) 
and a pile-up rate four times higher than the nominal value~($\mu=0.4$).
This posed a strong challenge to the trigger plus offline data processing 
and the experiment has demonstrated the ability to successfully cope
with these extreme running conditions.~Its overall data taking efficiency 
during the period 2010--2012 exceeded $90\%$ as can be seen in Fig.~\ref{fig:LHCbEff}.
To moderate the instantaneous luminosity at the LHCb collision point 
with respect to the LHC nominal interaction rate in 2011 and 2012, 
the accelerator was continuously displacing both proton beams to reduce 
the area of interactions where the proton bunches cross through each other.
Due to this technique, called luminosity levelling, the interaction rate 
at the LHCb collision point was kept at a roughly constant value 
throughout an entire LHC fill.

\section{Selected physics results}
\label{physics}

Using large samples of $pp$ collision data accumulated in 2010--2012, 
the LHCb collaboration has performed a series of measurements
that provides a sensitive test of the Standard Model
delivering valuable input to the existing theories.
Some of these are summarised hereafter. 

\subsection{Rare B hadron decays}

In the Standard Model, the rare leptonic and semileptonic decays 
$B^{0}_{(s)}\to\mu^{+}\mu^{-}$  and $B^{0}\to K^{*0}\mu^{+}\mu^{-}$ 
are highly suppressed FCNC processes mediated by electroweak box and penguin type diagrams. 
These decays have an enhanced sensitivity to physics 
beyond the Standard Model, as new heavy particles may enter in 
diagrams, which compete with the Standard Model processes, strongly affecting 
branching fractions or angular distributions of the daughter particles.
Looking for evidence of New Physics, the LHCb collaboration 
has recently studied these processes with currently available data.  
It should be noted that the outcome of these analyses imposes strong constraints 
on supersymmetric models~\cite{Implication}.

\subsubsection{ $B^{0}_{(s)}\to\mu^{+}\mu^{-}$ }

A recent search for the rare decays $B^{0}_{s}\to\mu^{+}\mu^{-}$
and $B^{0}\to\mu^{+}\mu^{-}$ is conducted with $1.0~{\rm fb}^{-1}$ and $1.1~{\rm fb}^{-1}$ 
of $pp$ collision data recorded at $\sqrt{s}=7$~TeV and $\sqrt{s}=8$~TeV, respectively~\cite{RareB1}. 
Signal candidates are selected among the triggered events 
by requiring two high quality muon candidates originating 
from a common vertex, which is well displaced with respect to any $pp$ interaction vertex, 
and having an invariant mass close to the $B^{0}_{(s)}$ nominal mass.
A multivariate selection, based on boosted decision trees~\cite{BDT},  
is applied at the next stage. It rigorously suppresses 
the background contamination retaining the majority of signal 
as determined from simulation. 
\begin{figure}[b!]
\centering
\resizebox{3.3in}{!}{
\rotatebox{0}{
\includegraphics{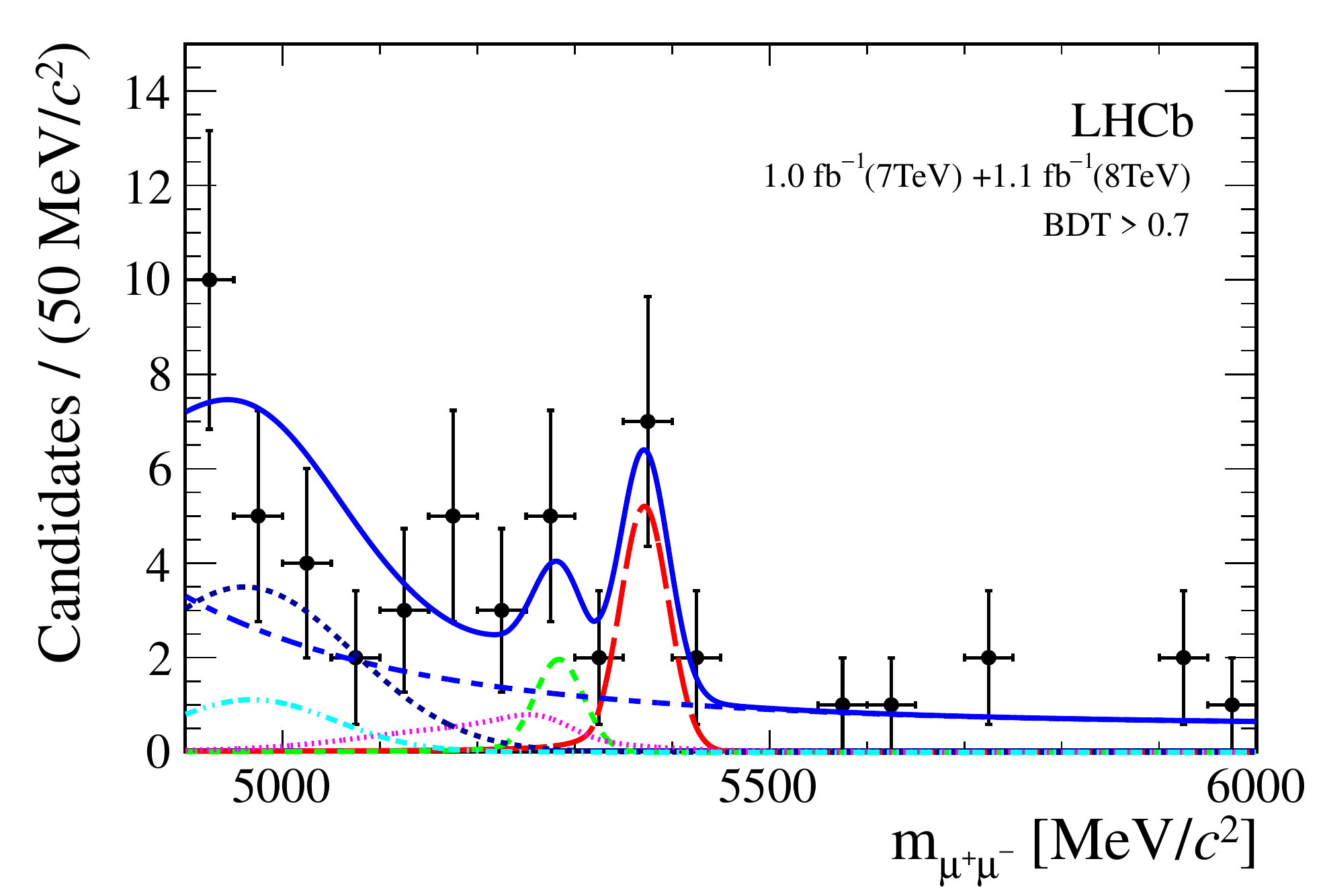}}}
\caption{Invariant mass distribution of the selected $B^{0}_{s}\to\mu^{+}\mu^{-}$ candidates. 
The result of the fit is illustrated (blue solid line) with different components:
$B^{0}_{s}\to\mu^{+}\mu^{-}$ (red long-dashed line), $B^{0}\to\mu^{+}\mu^{-}$ (green medium-dashed line), 
combinatorial background (blue medium-dashed line) and contamination from the other $B$ hadron decays (remaining lines).  
}
\label{fig:BsMuMu} 
\end{figure}

Taking into account the measured normalisation factors and assuming 
the Standard Model branching fractions, the combined 2011 and 2012 data sample 
is expected to contain about 14.1 $B^{0}_{s}\to\mu^{+}\mu^{-}$ and 1.7 
$B^{0}\to\mu^{+}\mu^{-}$ decays. 
Figure~\ref{fig:BsMuMu} illustrates the resulting invariant mass 
distribution of the selected $B^{0}_{s}\to\mu^{+}\mu^{-}$ candidates, 
where the signal shape is described by a Crystal Ball function.
The central peak values for the $B^{0}_{s}$ and $B^{0}$ mesons are obtained 
from the control channels $B^{0}_{s}\to K^{+}K^{-}$, $B^{0}\to K^{+}\pi^{-}$ and 
$B^{0}\to\pi^{+}\pi^{-}$. The resolutions are estimated by combining the results obtained 
with a power-law interpolation between the measured resolutions of charmonium and bottomonium resonances 
decaying into two muons with those obtained with a fit of the mass distributions of control channels.
%
%
The data in the $B^{0}_{s}$ search region show an excess of events 
with respect to the background expectation with a statistical 
significance of 3.5 standard deviations.~It implies that
such an excess of $B^{0}_{s}\to\mu^{+}\mu^{-}$ candidates 
is induced by the signal with probability of $1-5.3 \times 10^{-4}=0.99947$. 
By applying a maximum-likelihood fit to the data, the branching fraction 
is estimated to be ${\cal B}(B^{0}_{s}\to\mu^{+}\mu^{-}) = 3.2^{+1.5}_{-1.2} \times 10^{-9}$, 
where the uncertainty is fully dominated by the statistical component.
This is the first evidence for the decay $B^{0}_{s}\to\mu^{+}\mu^{-}$ and 
the measured branching fraction is in agreement with the Standard Model prediction.
In the $B^{0}$ search region, the data are consistent with the background expectation.
This result sets currently the world's best upper limit of 
${\cal B}(B^{0}\to\mu^{+}\mu^{-}) < 9.4 \times 10^{-10}$ at $95\%$ confidence level.

\subsubsection{ $B^{0}\to K^{*0}\mu^{+}\mu^{-}$ }
\begin{figure}[b!]
\centering
\resizebox{3.2in}{!}{
\rotatebox{0}{
\includegraphics{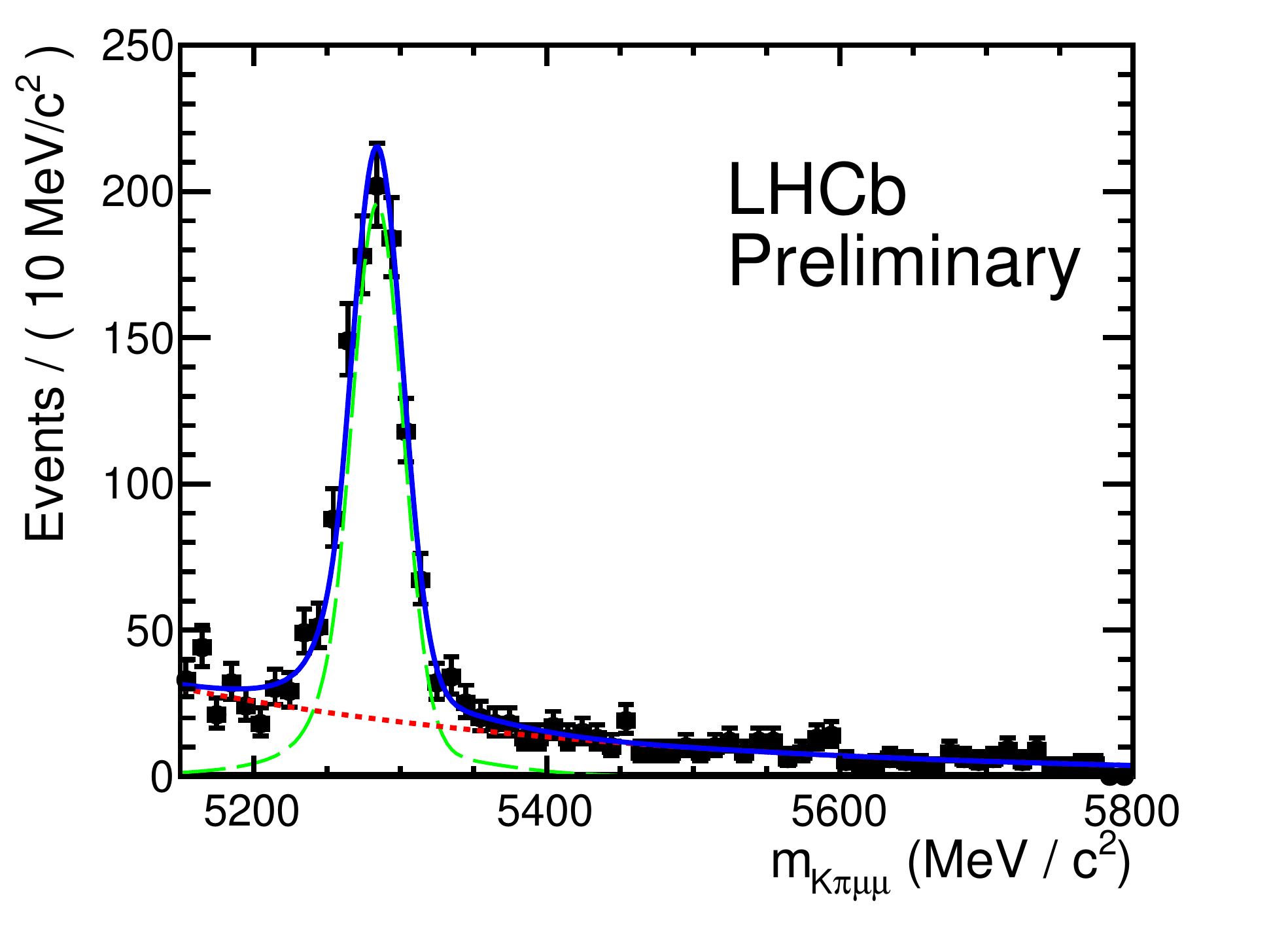}}}
\caption{Invariant mass distribution of the selected $B^{0}\to K^{*0}\mu^{+}\mu^{-}$ candidates
in the range $4m_{\mu}^{2}<q^{2}<19~{\rm GeV^{2}}/c^{4}$. 
The result of the fit is illustrated with the blue solid line, 
while the signal and background components are shown with the green long-dashed and red dashed lines, respectively.  
}
\label{fig:BdMuMuKst1} 
\end{figure}
Another FCNC decay of high interest, $B^{0}\to K^{*0}\mu^{+}\mu^{-}$, 
was recently studied with 2011 data sample corresponding to an integrated luminosity 
of $1.0~{\rm fb}^{-1}$~\cite{RareB2}. In this analysis, the differential branching fraction 
and a series of angular observables are investigated as a function 
of dimuon invariant mass squared, $q^{2}$. 
\begin{figure}[t!]
\centering
\resizebox{3.2in}{!}{
\rotatebox{0}{
\includegraphics{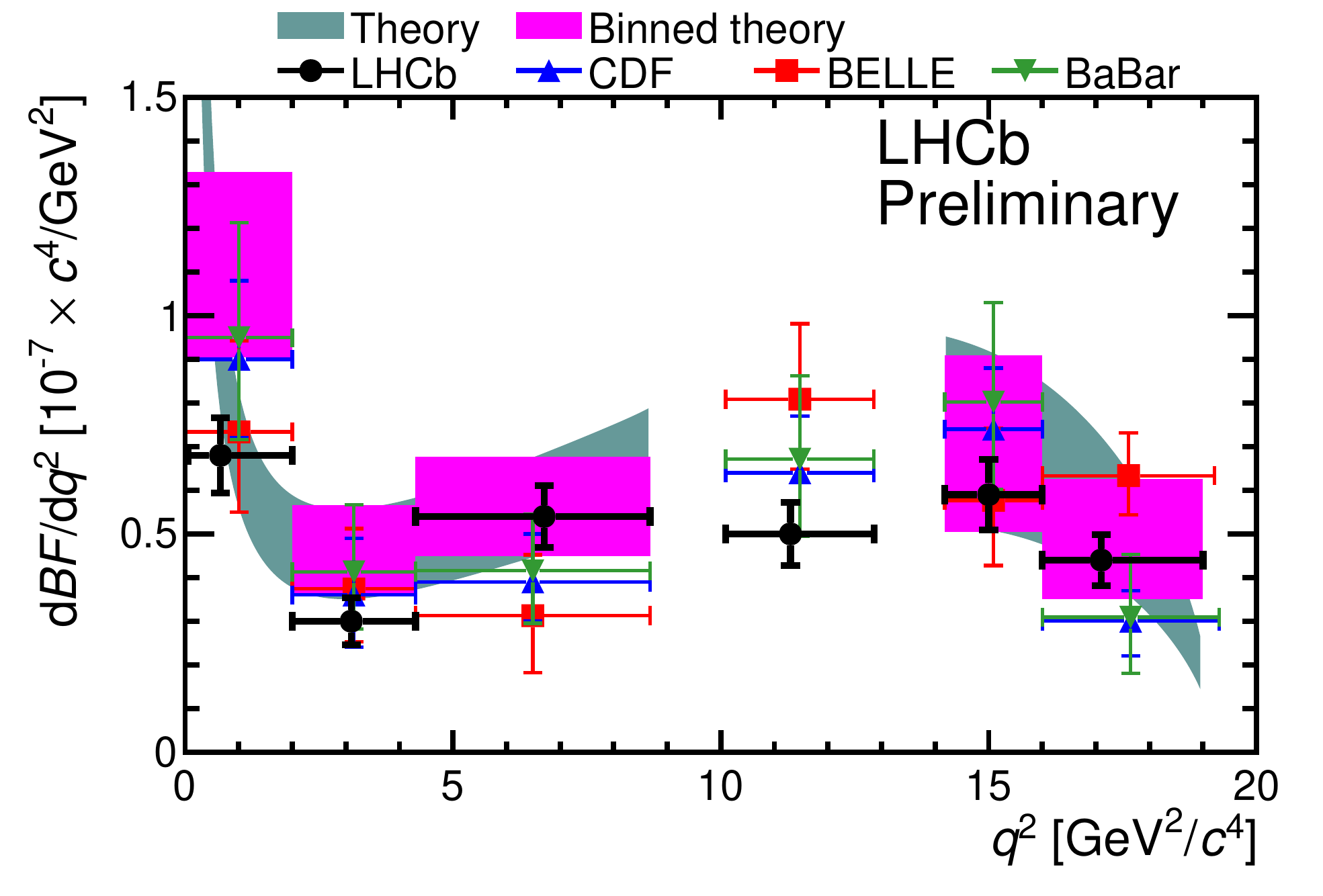}}}
\caption{
Differential branching fraction of the $B^{0}\to K^{*0}\mu^{+}\mu^{-}$ decay as a function of $q^{2}$.
Points include both statistical and systematic uncertainties. The BABAR~\cite{babar}, Belle~\cite{belle} 
and CDF~\cite{cdf} measurements are shown too. 
The theory predictions are described in Ref.~\cite{RareB2pred}.
}
\label{fig:BdMuMuKst2} 
\end{figure}
\begin{figure}[h!]
\centering
\resizebox{3.2in}{!}{
\includegraphics{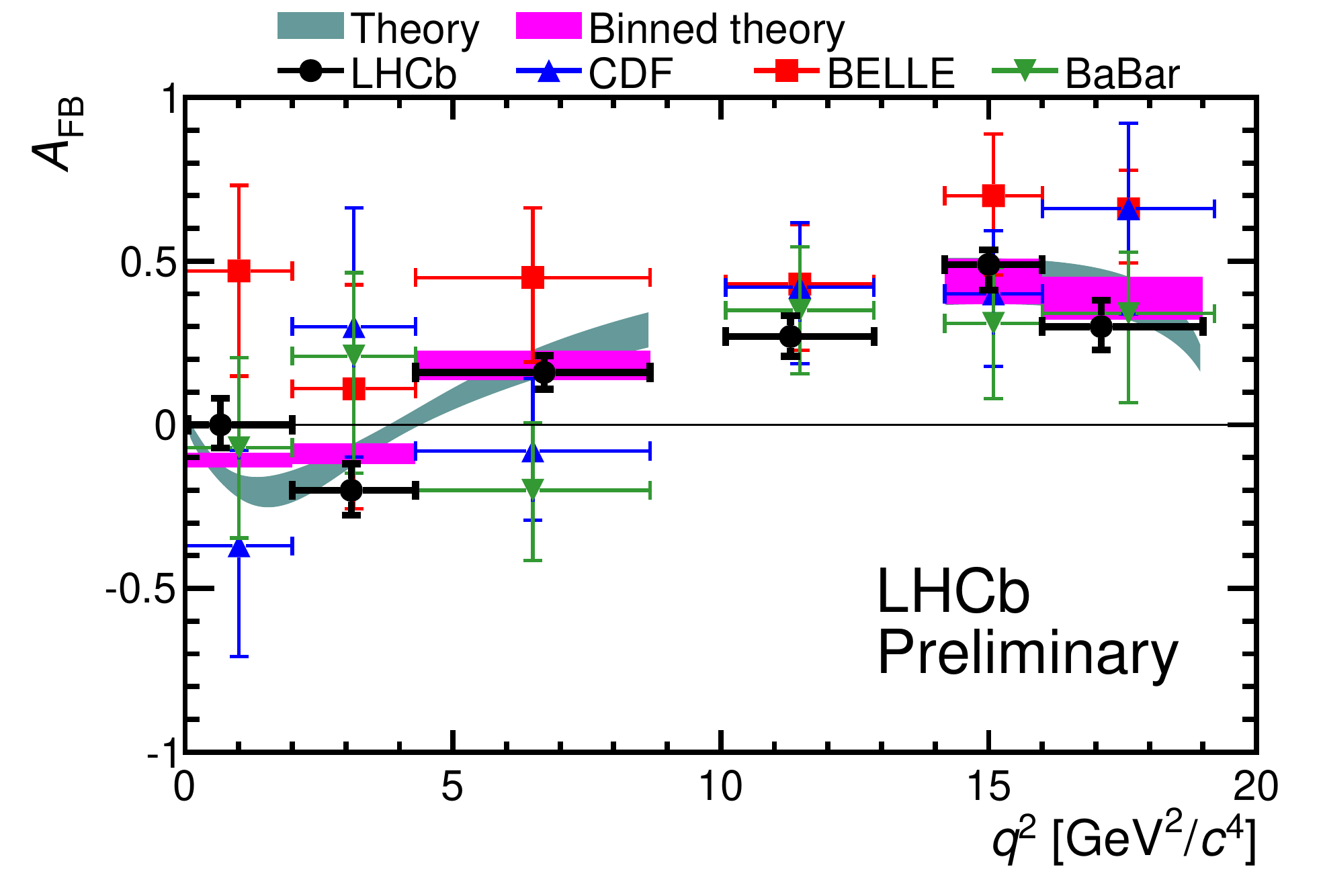}}
\newline
\resizebox{3.2in}{!}{
\includegraphics{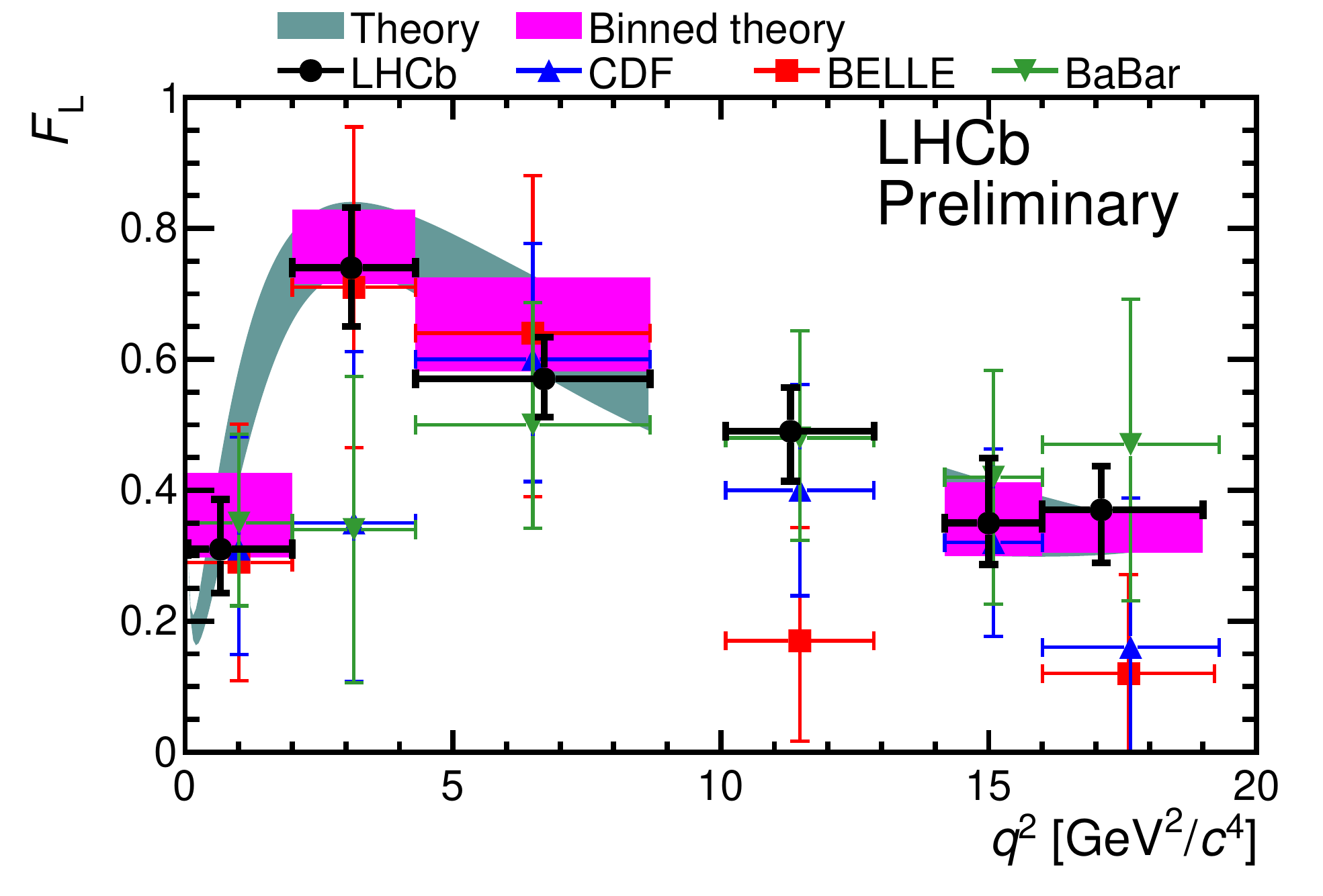}}
\caption{Forward-backward asymmetry of the dimuon system, $A_{\rm FB}$,~(top) and fraction of longitudinal polarisation 
of the $K^{*0}$ meson, $F_{\rm L}$,~(bottom) as a function of $q^{2}$. Points include both statistical 
and systematic uncertainties. The BABAR~\cite{babar}, Belle~\cite{belle}
and CDF~\cite{cdf} measurements are shown too.
The theory predictions are described in Ref.~\cite{RareB2pred}. }
\label{fig:BdMuMuKst34} 
\end{figure}
To select signal candidates among the triggered events a loose selection, 
based on the $B$ hadron decay time and daughter impact parameters, 
is applied at the first stage. A tight multivariate selection, using a boosted decision tree, 
is employed afterwards to suppress the remaining background. 
It allows to achieve a signal-to-background ratio in a $100~{\rm MeV}/c^2$ window
around the reconstructed $B$ mass of about three-to-one. 
Specific peaking backgrounds from $B$ hadron decay channels with similar topology 
are suppressed using tight particle identification requirements and excluding 
the regions of dimuon mass around the $c\bar{c}$ resonances.
The resulting invariant mass distribution of signal candidates 
in the range $4m_{\mu}^{2}<q^{2}<19~{\rm GeV^{2}}/c^{4}$ 
is shown in Fig.~\ref{fig:BdMuMuKst1}, while Fig.~\ref{fig:BdMuMuKst2} illustrates 
the observed differential branching fraction as a function of $q^{2}$.
The latter indicates that the measurements are in good agreement with the Standard Model predictions.
This is also the case for the angular observables of interest, some of which are shown in Fig.~\ref{fig:BdMuMuKst34}. 
A first measurement of the zero-crossing point of the forward-backward asymmetry of the dimuon system 
is additionally performed in this analysis. It is determined to be $q_{0}^{2}=(4.9_{-1.3}^{+1.1})~{\rm GeV^{2}}/c^{4}$, 
where the uncertainty is fully dominated by the statistical component.
The results obtained in this study are the most precise to date.

\subsection{$CP$-violating phase $\phi_{\rm s}$}

Using currently available data, the LHCb collaboration has performed 
high-precision studies of $CP$ asymmetry in different $B$ hadron decay 
channels~\cite{CPb1,CPb2,CPb3,CPb4,CPb5,CPb6,CPb7,CPb8,CPb9,CPb10,CPb11}.
One of the most interesting analyses was conducted to measure the mixing-induced 
$CP$-violating phase $\phi_{\rm s}$. It can be determined without theoretical 
uncertainties using $B_{s}^{0}$ decays proceeding via the $b \to c\bar{c}s$ 
transition by measuring the time-dependent $CP$ asymmetry, which arises 
in this process due to the interference between the amplitudes of the transition 
with and without oscillation. In the Standard Model, the $\phi_{\rm s}$ phase 
is small and accurately predicted, while it may be significantly 
enlarged by New Physics. As a result, its measurement provides 
a sensitive test of the Standard Model.

For accurate measurement of the $\phi_{\rm s}$ phase 
the LHCb collaboration has performed a series of independent analyses 
using the following $B_{s}^{0}$ decay channels 
which occur via the $b \to c\bar{c}s$ transition:  
$B_{s}^{0}\to J/\psi f_{0}(980)$~\cite{Phis1}, 
$B_{s}^{0}\to J/\psi \phi$~\cite{Phis2} 
and  
$B_{s}^{0}\to J/\psi \pi^{+}\pi^{-}$~\cite{Phis3}. 
In these studies, signal candidates are selected 
exploiting typical features of $B$ hadron decays
(long lifetime of $B$ hadrons, daughter particles with high $p_{\rm T}$) 
and imposing additional requirements on intermediate resonances and particle identification.
The flavour tagging is based on properties of the decay of the other $B$ hadron 
in the event and has an efficiency times dilution-squared of about $2\%$.  
The selected signal events are found to have an effective decay time 
resolution of better than 50~fs. The $\phi_{\rm s}$ phase is determined using 
a time-dependent fit to the data. The results obtained for 
the $B_{s}^{0}\to J/\psi \phi$ and $B_{s}^{0}\to J/\psi \pi^{+}\pi^{-}$ decays 
are combined by performing a simultaneous fit to the data. 
It yields a value of $\phi_{\rm s}=-0.002\pm0.083\pm0.027$~rad, where 
the first uncertainty is statistical and the second is systematic~\cite{Phis2}.
This result is the most precise to date and is consistent with 
the Standard Model expectation of $-0.0363^{+0.0016}_{-0.0015}$~rad.

\subsection{Evidence for $CP$ violation in the charm sector}

In addition to a rich $B$ physics programme, 
the LHCb collaboration aims to carry out high-precision measurements 
of all the key observables also in the charm sector.
The latter is feasible due to the high production rate of open charm 
and excellent detector performance allowing the selection of high purity samples 
of hadronic and muonic $D$ meson decays. Studies of $D$ mesons offer, in particular, 
a unique opportunity to access up-type quarks in FCNC processes. 
The primary goal of the experiment in this area of research is 
to explore $CP$ violation and $D^{0}-\bar{D}^{0}$ mixing along with 
a search for rare $D$ meson decays.

Using currently available data, LHCb has carried out various measurements in 
the charm sector~\cite{charm1,charm2,charm3,charm4,charm5,charm6,charm7,charm8,charm9,charm10,charm11,charm12}. 
One of the most interesting results was obtained by searching 
time-integrated $CP$ violation in $D^{0} \to h^{+}h^{-}~(h=K, \pi)$ 
decays using $0.62~{\rm fb}^{-1}$ of data recorded in 2011~\cite{CPcharm}. 
In this analysis, the difference in $CP$ asymmetry between 
$D^{0} \to K^{+}K^{-}$ and $D^{0} \to \pi^{+}\pi^{-}$ decays 
is measured to be $\Delta A_{\rm CP} = (-0.82\pm0.21\pm0.11)\%$, where 
the first uncertainty is statistical and the second is systematic.
This result differs from the hypothesis of $CP$ conservation 
by 3.5 standard deviations and thus, can be considered as 
the first evidence for $CP$ violation in the charm sector.

\subsection{Forward energy flow and charged particle multiplicities}

Besides an intensive heavy flavour physics programme, 
the LHCb experiment has performed important studies  
of QCD and electroweak processes in a unique kinematic range, 
which deliver valuable input to the knowledge of the 
parton density functions, underlying event activity, low Bjorken-x QCD dynamics 
and exclusive processes~\cite{Ks900,phi,V0ratio,HadronRatio,EP,EWboson,DY,Ztautau,Zee}.
Furthermore, some of these analyses have direct relevance 
to cosmic-ray and astroparticle physics.
In particular, measurements of forward particle production 
in high-energy hadron-hadron collisions give an opportunity to impose 
strong constraints on cosmic-ray interaction models 
which are widely used in extensive air shower simulations. 
In these models, the primary particle production is dominated 
by forward and soft QCD interactions, which are described  
by parameters constrained mainly using previous collider data. 
At LHC collision energies, a more reliable determination of the 
cosmic-ray energy and composition becomes possible~\cite{Enterria:2011}.

The first analysis where the LHCb data are compared to predictions 
given by cosmic-ray interaction models is the measurement of the energy flow 
$1/N_{\rm int}~dE_{\rm total}/d\eta$ created in $pp$ collisions at $\sqrt{s}=7$~TeV
within the pseudorapidity range $1.9<\eta<4.9$~\cite{EFpaper}. 
At large values of pseudorapidity this observable is expected 
to be directly sensitive to the amount of parton radiation 
and multi-parton interactions~\cite{PhysRevD.36.2019}.
The latter represent a predominant contribution to the soft component 
of a hadron-hadron collision, called the underlying event.
Its precise theoretical description still remains a challenge and 
this measurement is conducted to constrain the corresponding models. 
As described in Ref.~\cite{EFpaper}, the primary measurement is the energy flow carried 
by charged particles, while a data-constrained Monte Carlo estimate of the neutral component 
is used for the measurement of the total energy flow.
\begin{figure*}[t!]
  \centering
  \subfigure{\includegraphics[scale=0.31]{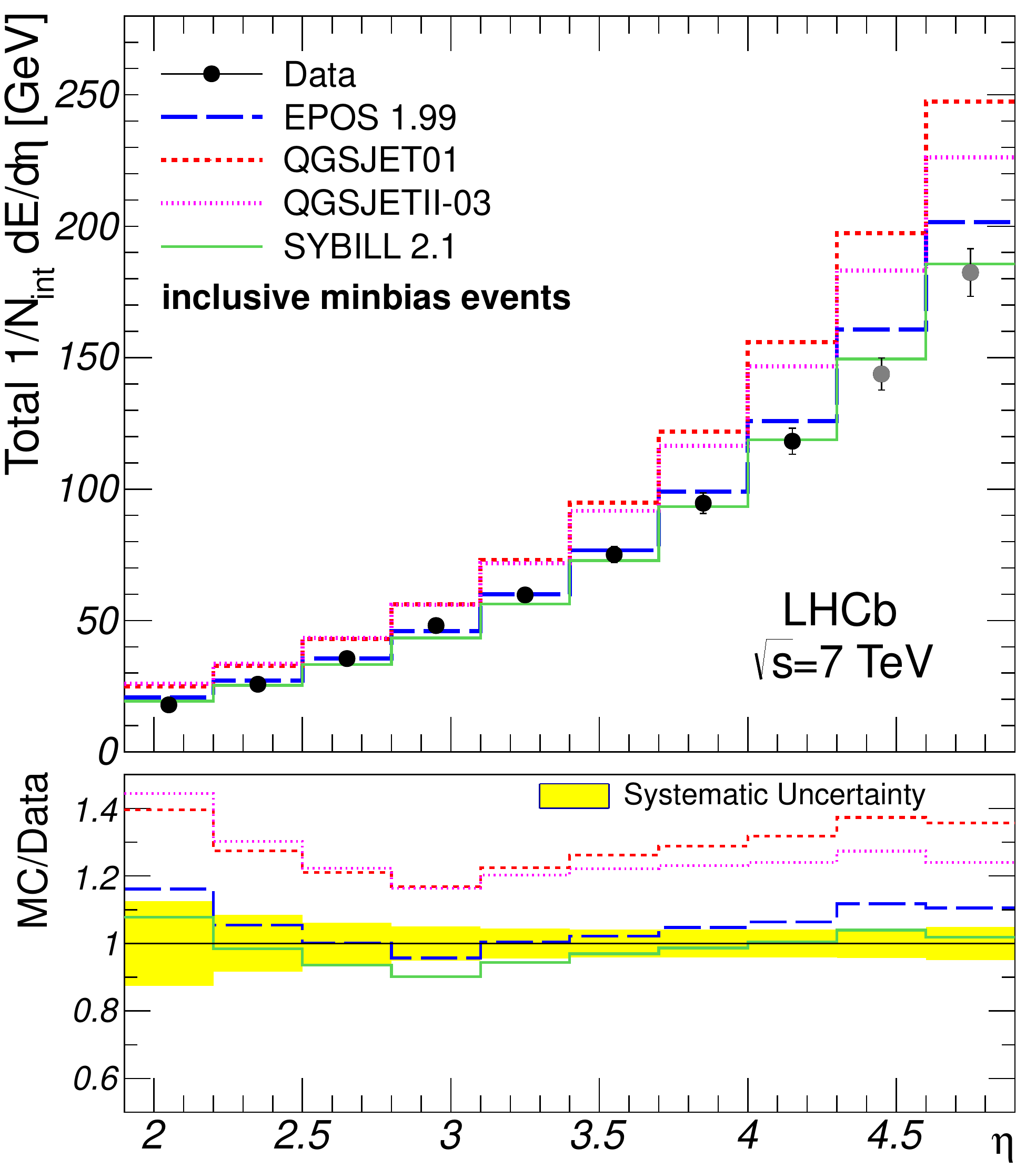}}\quad
  \subfigure{\includegraphics[scale=0.31]{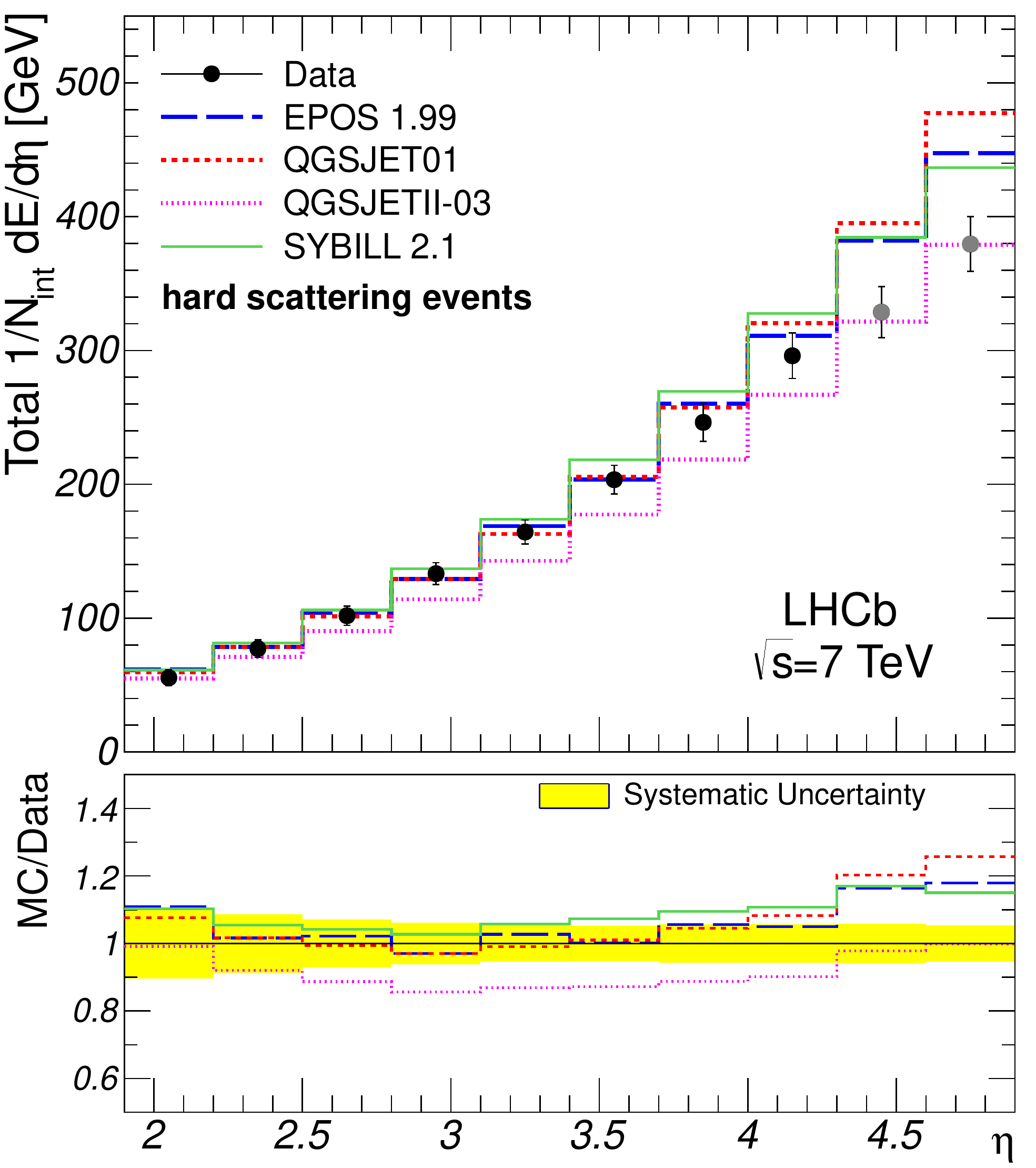}}
  \subfigure{\includegraphics[scale=0.31]{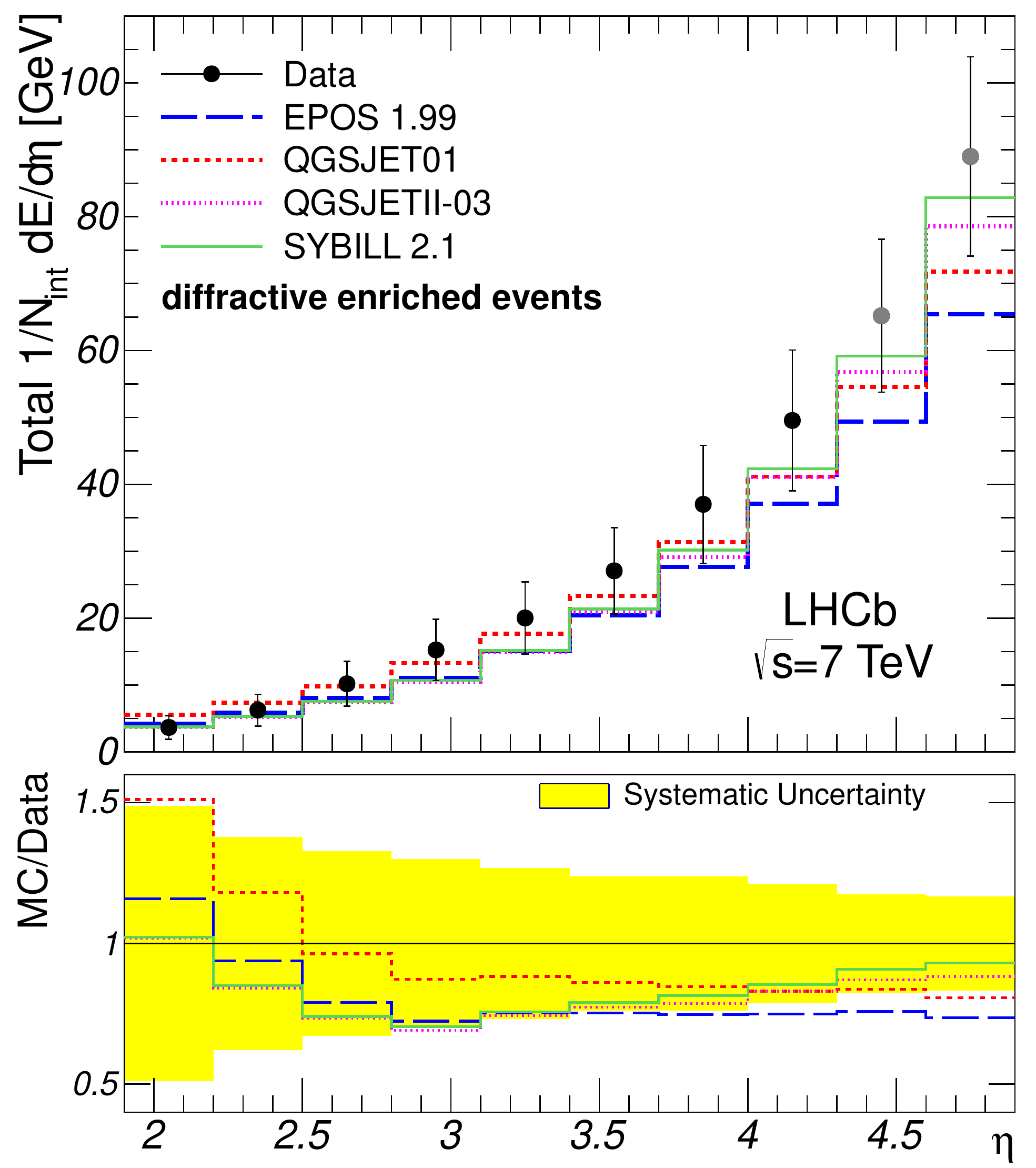}}\quad
  \subfigure{\includegraphics[scale=0.31]{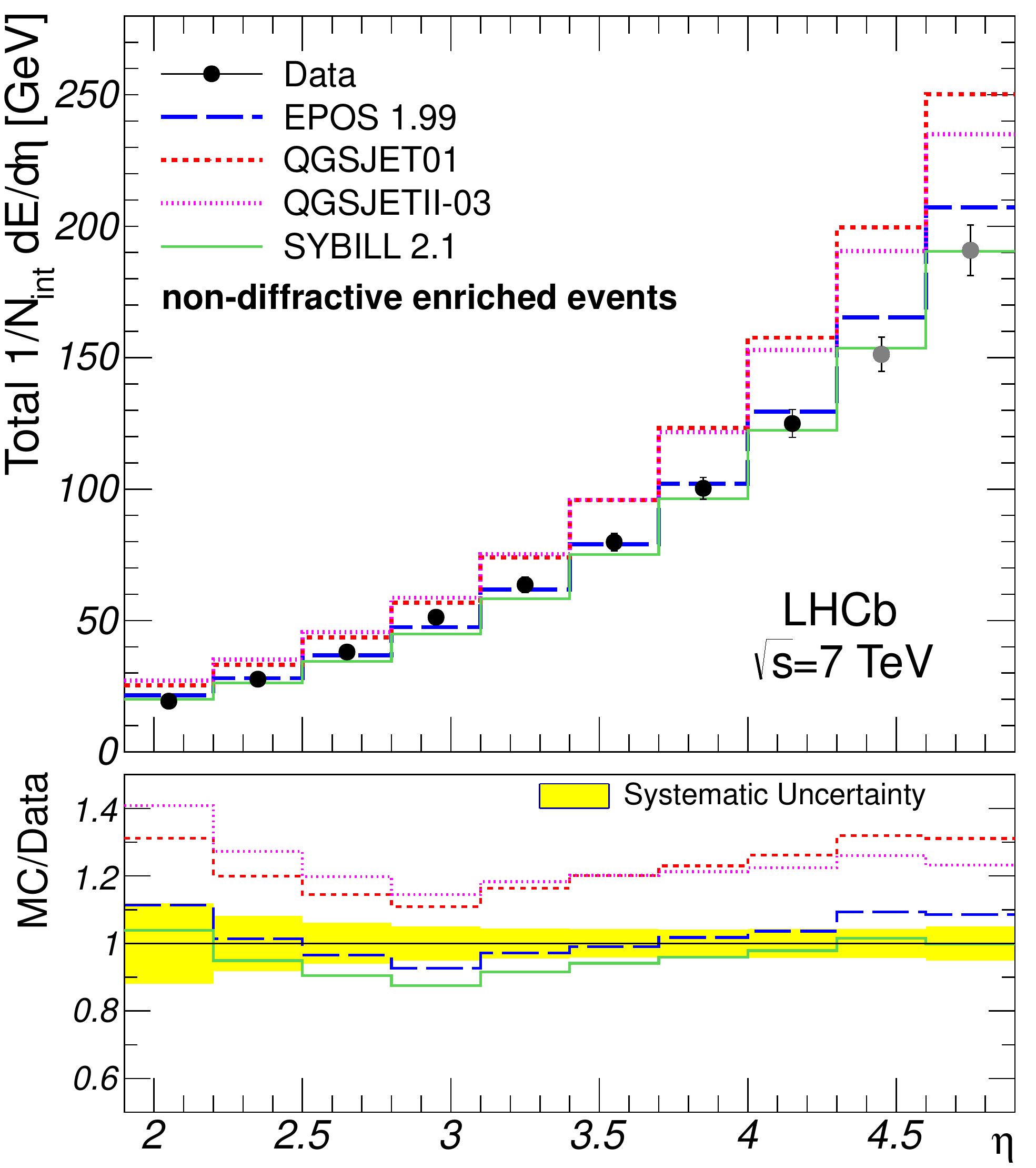}}
  \caption{Total energy flow as a function of $\eta$ for all event classes as indicated in the figures. 
          The corrected measurements are given by points with error bars, while the predictions by the cosmic-ray 
          interaction models are shown as histograms. The comparison of the data against {\sc Pythia}-based predictions 
          can be found in Ref.~\cite{EFpaper}. 
          The error bars represent the systematic uncertainties, which are highly correlated between the bins. }
 \label{fig:EF}	  
\end{figure*}
To probe various aspects of multi-particle 
production in hadron-hadron collisions, 
the measurements are performed for inclusive minimum-bias (containing 
at least one track with $p>2$~GeV/c in $1.9<\eta<4.9$), 
hard scattering (having at least one track with $p_{\rm T}>3$~GeV/c in $1.9<\eta<4.9$), 
diffractive, and non-diffractive enriched interactions. 
The last two event types were selected among the inclusive minimum-bias interactions 
requiring the absence or presence of at least one backward track reconstructed by 
the VELO in $-3.5<\eta<-1.5$, respectively. 
In addition to cosmic-ray interaction models, experimental results are compared 
to predictions given by the {\sc Pythia}-based event generators~\cite{Skands:2009zm,Clemencic:LHCbMC,P8},   
which model the underlying event activity in different ways.
Though the evolution of the energy flow as a function of $\eta$ 
is reasonably well reproduced by the Monte Carlo event generators, 
none of the models used in this analysis are able to describe the energy flow measurements 
for all event classes that have been studied. 
The majority of the {\sc Pythia} tunes underestimate the measurements at large $\eta$, 
while most of the cosmic-ray interaction models overestimate them as can be seen in Fig.~\ref{fig:EF}.
The energy flow is found to increase with the momentum transfer in an underlying $pp$ inelastic interaction.

Another LHCb analysis constraining the underlying event models 
being relevant for extensive air shower simulations is the measurement 
of charged particle production in $pp$ collisions at $\sqrt{s}=7$~TeV~\cite{mult}.
In this study, the multiplicity and density distributions 
of charged particle tracks originating from the primary interaction 
are investigated as a function of $\eta$ within the ranges 
$2.0<\eta<4.5$ and $-2.5<\eta<-2.0$ exploiting high detection 
efficiency of the VELO in these phase space regions.
The measurements are carried out for events recorded 
with a minimum bias trigger requiring the presence 
of at least one reconstructed track in the VELO 
and for events containing at least one track with $p_{\rm T}>1~{\rm GeV}/c$.
The results are compared to predictions given by several 
{\sc Pythia}-based event generators~\cite{Skands:2009zm,Clemencic:LHCbMC,P8}. 
None of these are able to describe fully the charged particle multiplicity 
and density distributions as a function of $\eta$. 
The considered {\sc Pythia} tunes are found 
to underestimate the charged particle production 
in the phase space regions of the measurements.

\section{Summary}
\label{summary}

The LHCb spectrometer possesses excellent tracking and vertexing performance, 
robust particle identification and flexible trigger system.
The experiment has demonstrated the ability to take data of high quality
successfully coping with extreme running conditions. 
Using data collected in the years 2010--2012, the LHCb collaboration 
has performed the world's most precise measurements of various important 
physics observables and has observed for the first time different rare processes 
in the heavy flavour sector. In addition, the experiment carries out 
important measurements of QCD and electroweak processes in a unique kinematic range.
The physics potential of the experiment is expected to be fully exploited
at higher LHC collision energy and interaction rate with an upgraded detector
possessing enhanced trigger capabilities~\cite{frameworkTDR}. This would allow an order of
magnitude more data to be collected and improving sensitivity
to many important physics processes.

\end{document}